\documentclass[%
 reprint,
superscriptaddress,
showpacs,preprintnumbers,
nofootinbib,
 amsmath,amssymb,onecolumn,notitlepage,
 bm,
 aps,
]{revtex4-1}

\usepackage[dvipdfmx]{graphicx}
\usepackage{here}
\usepackage{amssymb}
\usepackage{comment}
\usepackage{physics}
\usepackage{epstopdf}
\usepackage{dcolumn}
\usepackage{booktabs}
\usepackage{bm}
\usepackage[dvipsnames]{xcolor}
\usepackage[caption=false]{subfig}
\usepackage{mhchem}

\usepackage{tensor}
\usepackage[dvipsnames]{xcolor}
\usepackage{hyperref}
\hypersetup{
colorlinks=true,
linkcolor=blue,
citecolor=BlueViolet,
urlcolor=Aquamarine
}

\usepackage[normalem]{ulem}

\usepackage[export]{adjustbox}% http://ctan.org/pkg/adjustbox

\providecommand{\sorthelp}[1]{}
\begin{document}

%\preprint{APS/123-Q ED}

\title{
Mapping the galaxy-halo connection to the galaxy bias:\\
implication to the HOD-informed prior
}

\newcommand{\kek}{Theory Center, Institute of Particle and Nuclear Studies,
High Energy Accelerator Research Organization (KEK), Tsukuba, Ibaraki 305-0801, Japan}

\author{Kazuyuki Akitsu}
\email{kakitsu@post.kek.jp}
\affiliation{\kek}

\date{\today}% It is always \today, today,
    % but any date may be explicitly specified
    
\begin{abstract}
The galaxy bias parameters are crucial for modeling the large-scale structure in cosmology, yet uncertainties in these parameters often degrade the precision of cosmological constraints.
In this work, we investigate how different Halo Occupation Distribution (HOD) models impact the priors of the galaxy bias parameters, particularly focusing on quadratic bias parameters.
We generate galaxy mock catalogs using various HOD models, including a standard model and one incorporating halo concentration dependence to account for assembly bias, and measure the galaxy bias parameters with high precision using the quadratic field method.
We show that the inclusion of assembly bias associated to halo concentration could significantly impact the distributions of quadratic galaxy bias parameters, especially $b_2$.
Our findings suggest that accounting for assembly bias or other galaxy-halo connection models is important for obtaining accurate priors on the galaxy bias parameters, 
thereby improving the robustness of cosmological analyses with galaxy clustering.
\end{abstract}

\maketitle

\section{Introduction}
The three-dimensional galaxy clustering provides a wealth of cosmological information, such as the initial conditions of the universe, the nature of dark energy, and the properties of dark matter.
In fact, the galaxy clustering has been used to constrain the cosmological parameters with BOSS and eBOSS surveys~\cite{BOSS:2016wmc,Gil-Marin:2016wya,eBOSS:2020yzd,Ivanov:2019pdj,DAmico:2019fhj,Colas:2019ret,Philcox:2020vvt,Kobayashi:2021oud,Chen:2021wdi,Zhang:2021yna}.
Given the success of these surveys,
several projects are currently underway or planned, 
including the Dark Energy Spectroscopic Instrument (DESI)~\cite{DESI:2013agm,DESI:2024uvr,DESI:2024mwx}, 
the Euclid mission~\cite{Euclid:2024yrr}, 
the Prime Focus Spectrograph (PFS)~\cite{PFSTeam:2012fqu}, 
and the Nancy Grace Roman Space Telescope~\cite{Spergel:2015sza}. 
By analyzing the clustering of galaxies with unprecedented precision provided by these surveys, 
we will be able to obtain stringent constraints on the cosmological parameters, giving a deeper insight into the nature of the universe.

However, the observed clustering pattern of galaxies is affected by both the cosmological and astrophysical effects.
To account for the latter effect while getting the unbiased information on cosmology, 
usually we introduce a number of nuisance parameters, including the galaxy bias parameters, which are the key ingredients in the analysis of the large-scale structure (LSS) data with the perturbative method~\cite{Desjacques:2016bnm,Bernardeau:2001qr,Fry:1992vr,McDonald:2006mx,McDonald:2009dh,Assassi:2014fva,Perko:2016puo}.
While the bias expansion is universal in the sense that it can be applied to any tracer of the LSS and is valid on large scales regardless of the details of the galaxy formation (i.e., UV physics), 
their specific values depend on the tracer of LSS.

Because we do not have a first-principle model for the galaxy formation, 
%the galaxy bias parameters are usually treated as free parameters in the analysis of the LSS data.
in practice, we have to put wide priors on the galaxy bias parameters in the analysis of the LSS data, which leads to the degradation of the constraining power of the cosmological parameters
due to the degeneracies between the galaxy bias parameters and the cosmological parameters.
Furthermore, wide priors on the galaxy bias parameters could also lead to the biased estimation of the cosmological parameters, known as the prior volume effect~\cite{Simon:2022lde,Carrilho:2022mon,Gomez-Valent:2022hkb,DAmico:2022osl,Holm:2023laa}.
Hence, it would be helpful to have a tight prior on the galaxy bias parameters to get unbiased and improved constrains on the cosmological parameters.

Recently several works employed the Halo Occupation Distribution (HOD) model as the galaxy-halo connection to get the prior on the galaxy bias parameters~\cite{Ivanov:2024xgb,Cabass:2024wob,Ivanov:2024hgq,Zhang:2024thl}.
The HOD is a phenomenological model that describes the probability that a halo of a given mass contains a certain number of galaxies~\cite{Peacock:2000qk,Cooray:2002dia,Berlind:2002rn,Zheng:2004id}.
Since populating galaxies onto halos based on the HOD is computationally much cheaper than the galaxy-formation simulation,
we can generate a large number of galaxy mock catalogs with different HOD parameters, which can be used to measure the galaxy bias parameters.
Combined with the conservative prior on the HOD parameters, we could obtain informative priors on the galaxy bias parameters, which can be used to improve the constraining power of the cosmological parameters and resolve the issues associated with the prior.
On the other hand, this implies that the obtained prior on the galaxy bias parameters depends on the choice of the HOD model, the range of the HOD parameters, the method to measure the galaxy bias parameters.
Therefore, it is important to understand the impact of these choices on the prior on the galaxy bias parameters.

In this paper, we aim to investigate the impact of the HOD model on the prior on the galaxy bias parameters.
While the previous studies used the {\tt ABACUSHOD} code~\cite{Yuan:2021izi}, we consider yet another HOD model in populating galaxies onto halos in addition to the standard one.
In particular, we consider the HOD model that depends on halo concentration \cite{Kobayashi:2019jrn}, motivated by the fact that the halo bias depends not only on the halo mass but also on halo concentration, known as the assembly bais~\cite{Jing:2006ey,Wechsler:2005gb,Gao:2006qz,Dalal:2008zd}.
In order to measure the galaxy bias parameters accurately and quickly, we make use of the quadratic field method~\cite{Schmittfull:2014tca,Lazeyras:2017hxw,Abidi:2018eyd,Akitsu:2023eqa}, which is based on the Eulerian field-level comparison.

The paper is organized as follows: Section \ref{sec:method} provides a description of the HOD model used in our study and the method we use to measure bias parameter precisely. 
We show and discuss the results in Section \ref{sec:result}. 
Finally, Section \ref{sec:discussion} concludes the paper and outlines potential directions for future research.
In Appendix \ref{app:halo_bias}, we compare our measurement of the halo bias parameters with the literature.

\section{Method}
\label{sec:method}
In this section, we describe the HOD models we use in this study in the first half and the method to measure the galaxy bias parameters from the galaxy mock catalogs in the latter half.

\subsection{Galaxy mock from HOD}
\label{subsec:HOD}
In general, it is still a challenging task to create a realistic galaxy mock, since the galaxy-formation simulation is computationally expensive while the gravity-only $N$-body simulation is cheaper.
The common tactic to make galaxy catalogs is to populate galaxies onto halos generated from gravity-only simulations, following a phenomenological halo-galaxy connection model.
The halo occupation distribution (HOD) is a popular empirical approach to populating galaxies onto dark-matter halos. 
Instead of modeling galaxy formation from the first principles, it provides a probabilistic framework for the halo-galaxy connection. 
The standard HOD model divides galaxies into central and satellite galaxies and specifies the mean number of each type of galaxies as a function of only host halo mass ($M$):
\begin{align}
    \langle N_c \rangle (M) & = \frac12 \left[ 1 + {\rm erf}\left(\frac{\log_{10}M - \log_{10}M_{\rm min}}{\sigma_{\log_{10}M}} \right) \right]\,\,,
    \label{eq:Nc}
    \\
    \langle N_s \rangle (M) & = \langle N_c \rangle (M) \left(\frac{M - \kappa M_{\rm min}}{M_1}\right)^\alpha\,\,,
    \label{eq:Ns}
\end{align}
where $\mathrm{erf}(x)$ is the error function, $M_{\rm min}$ and $\sigma_{\log_{10} M}$ determine the typical minimum mass cut and the softness of this cut, 
and $\kappa$, $M_1$ and $\alpha$ determine the profile of the mean number of satellite galaxies.
With these mean values, galaxies are assigned following a Bernoulli distribution for centrals and a Poisson distribution for satellites.
We populate satellite galaxies into halos that already host a central galaxy.
In order to take into account for model variations of HOD, we consider three different HOD models as follows:
\begin{itemize}
    \item Simplified HOD: Galaxies are populated only according to the Bernoulli distribution with the mean \eqref{eq:Nc} for both host and sub halos identified by \texttt{Rockstar}~\cite{Behroozi:2011ju}. 
    In this case, we only have the two parameters; $M_{\rm min}$ and $\sigma_{\log_{10}M}$ (see \cite{Nishimichi:2020tvu} for details).
    \item Standard HOD: Galaxies are populated only for host halos as described above.
    \item Standard HOD with concentration: This is based on the standard HOD procedure, but we add a dependence on halo concentration as follows, inspired by \cite{Kobayashi:2019jrn}:
    We first divide halos into 100 mass bin in logarithmic scale from $M = 10^{12} M_\odot/h$ to $M = 10^{16} M_\odot/h$. Then we create a ranked list of all halos ordered in descending order based on their concentration.
    Central galaxies are subsequently assigned to halos starting from the top of this list (i.e., the highest-concentration halos) in proportion to the number fraction specified by the central HOD,~\eqref{eq:Nc}.
    Satellite galaxies are assigned to halos in the same way as the standard.
    This method maintains the average HOD as a function of halo mass but modifies the clustering properties of galaxies by preferentially selecting host halos with higher concentrations.
\end{itemize}

We populate galaxies with \texttt{Rockstar} halos identified in eight independent realizations of the $N$-body simulation of each $1.5~{\rm Gpc}/h$ length and with $1536^3$ particles, employed in~\cite{Akitsu:2023eqa}.
For each type of the HOD models we produce 44,550 mock catalogs of galaxies, varying the HOD parameters in the following ranges:
\begin{align}
    \log_{10}M_{\rm min} \in & [12.4, 14.2]\,\,,
    \label{eq:HOD_range_Mmin}
    \\
    \sigma_{\log_{10}M} \in & [0.1, 1.0]\,\,,
    \label{eq:HOD_range_sigmaM}
    \\
    \kappa \in & [0.1, 1.0]\,\,,
    \label{eq:HOD_range_kappa}
    \\
    \log_{10} M_1 \in & [13.0, 15.0]\,\,,
    \label{eq:HOD_range_M1}
    \\
    \alpha \in & [0.0, 1.6]\,\,.
    \label{eq:HOD_range_alpha}
\end{align}
The HOD galaxy mocks with difference HOD parameter sets are generated at $z=0.38$, $0.51$ and $0.61$, corresponding to LOWZ, CMASS1 and CMASS2, respectively.

\subsection{Precise measurement of the bias parameters}
\label{subsec:bias_measurement}
Here we explain the method to precisely measure the galaxy bias parameters from the simulations.
Up to the quadratic order, the galaxy (or halo) density fields can be written as 
\begin{align}
    \delta_g({\bf x}) = & b_1 (\delta({\bf x}) + \delta^{(2)}({\bf x})) + \frac{b_2}{2}(\delta^2({\bf x}) - \sigma^2) + b_{{\cal G}_2} {\cal G}_2({\bf x})
    \\
    = &  b_1 (\delta({\bf x}) + \delta^{(2)}({\bf x})) + \frac{\tilde{b}_2}{2}(\delta^2({\bf x}) - \sigma^2) + b_{K^2} K^2({\bf x}),
\end{align}
where $\delta^{(2)}$ is the second-order density field and $K_2$ and ${\cal G}_2$ are tidal bias operators in the different basis:
\begin{align}
    {\cal G}_2({\bf x}) & =  \left(\frac{\partial_i\partial_j}{\partial^2}\delta({\bf x}) \right)^2 - \delta^2({\bf x}), \\
    K^2({\bf x}) & = K^2_{ij}({\bf x}), \hspace{0.2cm}\mathrm{with}\hspace{0.2cm}
    K_{ij}({\bf x}) = \left(\frac{\partial_i\partial_j}{\partial^2} -\frac13 \delta_{ij} \right)\delta({\bf x}) .
\end{align}
Note that the value of $b_2$ depends on the basis for the tidal operators and we distinguish them by the tilde, i.e., $b_2 = \tilde{b}_2 + \frac43 b_{{\cal G}_2}$ and $b_{K^2} = b_{{\cal G}_2}$. 
The second-order density can be expressed in terms of the quadratic bias operators and the displacement field $ \Psi_i = -(\partial_i/\partial^2) \delta$:
\begin{align}
    \delta^{(2)}({\bf x}) = & \delta^2({\bf x}) + \frac27 {\cal G}_2({\bf x}) - \Psi({\bf x})\cdot\nabla\delta({\bf x}) \\
     = & \frac{17}{21}\delta^2({\bf x}) + \frac 27 K^2({\bf x})  - \Psi({\bf x})\cdot\nabla\delta({\bf x}).
\end{align}

The common strategy to measure the quadratic bias parameters is to compare the theoretical template of the bispectrum with its measurements~\cite{Baldauf:2013hka}.
However, this method greatly suffers from sample variance. 
In particular, the tree-level bispectrum is valid only on large scales where the sample variance is severe.
In this paper, we instead take full advantage of the simulations, where we know the realization of the initial conditions, whose uncertainty in reality introduces the sample variance.
To do so, we take the cross-correlation between the galaxy density field with the bias operators constructed from the initial conditions;

\begin{align}
    \langle \delta_g  {\cal O}_j \rangle = \sum_i b_i \langle {\cal O}_i {\cal O}_j \rangle ,
    \label{eq:field_cross_corre}
\end{align}
where ${\cal O}_i = \{ \delta, \delta^2/2, {\cal G}_2, -\Psi\cdot\nabla\delta \}$ or $\{ \delta, \delta^2/2, K^2, -\Psi\cdot\nabla\delta \}$ and $b_i = \{b_1, b_2, b_{{\cal G}_2}, 1 \}$ or $\{b_1, \tilde{b}_2, b_{K^2}, 1 \}$, respectively.
The point is that we can directly measure all the correlators appeared on both sides in Eq.\eqref{eq:field_cross_corre} from the simulations and its initial conditions so that the sample variance are almost canceled out.
Specifically, we consider the following likelihood,
\begin{align}
    \chi^2 = \sum_i \sum_k^{k_{\rm max}} \frac{| \langle \delta_g  {\cal O}_i \rangle (k) - \sum_j b_j \langle {\cal O}_j {\cal O}_i \rangle(k) |^2}{ \sigma^2(\langle \delta_g  {\cal O}_i \rangle)(k))}
\end{align}
to measure bias parameters by running MCMC. We set $k_{\rm max} = 0.03 ~h/{\rm Mpc}$ for ${\cal O}_i = \delta$ and $k_{\rm max} = 0.08 ~h/{\rm Mpc}$ for other quadratic fields.
This method was first proposed in Ref.~\cite{Schmittfull:2014tca} and subsequently used to measure the cubic halo bias parameters in Refs.~\cite{Lazeyras:2017hxw, Abidi:2018eyd} and the quadratic shape bias in Ref.~\cite{Akitsu:2023eqa}.

We also filter out the high-$k$ modes of the initial conditions when constructing the relevant bias operators to obtain the ``renormalized bias'', which is defined as the low-$k$ limit of polyspectra~\cite{Assassi:2014fva, Desjacques:2016bnm}. We apply the Gaussian smoothing $\exp(-k^2R^2/2)$ with $R = 20~{\rm Mpc}/h$, following Refs.~\cite{Abidi:2018eyd, Akitsu:2023eqa}.
We confirm that $R=15 {\rm Mpc}/h$ gives consistent results, so our results are stable against different choice of the smoothing scale, as long as it is sufficiently large.
We also compare the measurement of the halo bias parameters with this method to the results in the literature, finding a great agreement, which is summarized in App.~\ref{app:halo_bias}.

\section{Result}
\label{sec:result}
In this section, we show the measurements of galaxy bias parameters from the HOD mocks and discuss its implications.

\subsection{Relation between the halo bias and the galaxy bias}
Before showing the measurements, here we discuss the basic properties of the quadratic bias parameter of HOD-galaxies to get insight on the quadratic galaxy bias parameters.

Given the halo mass function, $\dd n/\dd M$, the local number density of halos with mass $M$ is written as 
\begin{align}
    n_{\rm h}({\bf x};M) = \frac{\dd n}{\dd M}\left[ 1 + \delta_{\rm h}({\bf x}) \right].
\end{align}
Similarly, the local number density of galaxies can be expressed as 
\begin{align}
    n_{\rm g}({\bf x}) = \bar{n}_{\rm g} \left[ 1 + \delta_{\rm g}({\bf x}) \right],
\end{align}
where $\bar{n}_{\rm g}$ is the mean number density of galaxies, which is related to the halo mass function via the total HOD number $\langle N_g\rangle= \langle N_c\rangle + \langle N_s \rangle$ as
\begin{align}
    \bar{n}_{\rm g} = \int \dd M \frac{\dd n}{\dd M} \langle N_g \rangle.
\end{align}
Here we focus on the standard HOD.
Neglecting the spatial dependence of the HOD, this relation holds for the local number density as well,
\begin{align}
    n_{\rm g}({\bf x}) & = \int \dd M \frac{\dd n}{\dd M}[1 + \delta_{\rm h}({\bf x})] \langle N_g \rangle
    \nonumber \\
    & = \bar{n}_{\rm g}\left[1 + \frac{1}{\bar{n}_{\rm g}}  \int \dd M \frac{\dd n}{\dd M} \delta_{\rm h}({\bf x})\langle N_g \rangle  \right]
\end{align}
Combining the bias expansion, this give rise to the following relation between halo bias and galaxy bias:
\begin{align}
    b^{\rm g}_i =  \frac{1}{\bar{n}_{\rm g}}  \int \dd M \frac{\dd n}{\dd M} \langle N_g \rangle b_i^{\rm h}(M).
\end{align}
This means that the bias parameters of galaxies are obtained on weighted average over the halo bias parameters.
In other words, the galaxy bias parameters are given by the expectation over the probability $\propto \langle N_g \rangle \dd n/\dd M$.
This gives a basic understanding of the relation between the halo bias and the galaxy bias, which is useful to interpret the results of the galaxy bias parameters.

\subsection{The dependence of the quadratic galaxy bias on the HOD parameters}
\label{subsec:bias_HOD_fiducial}

\begin{figure}[tb]
    \centering
    \includegraphics[width=0.49\textwidth]{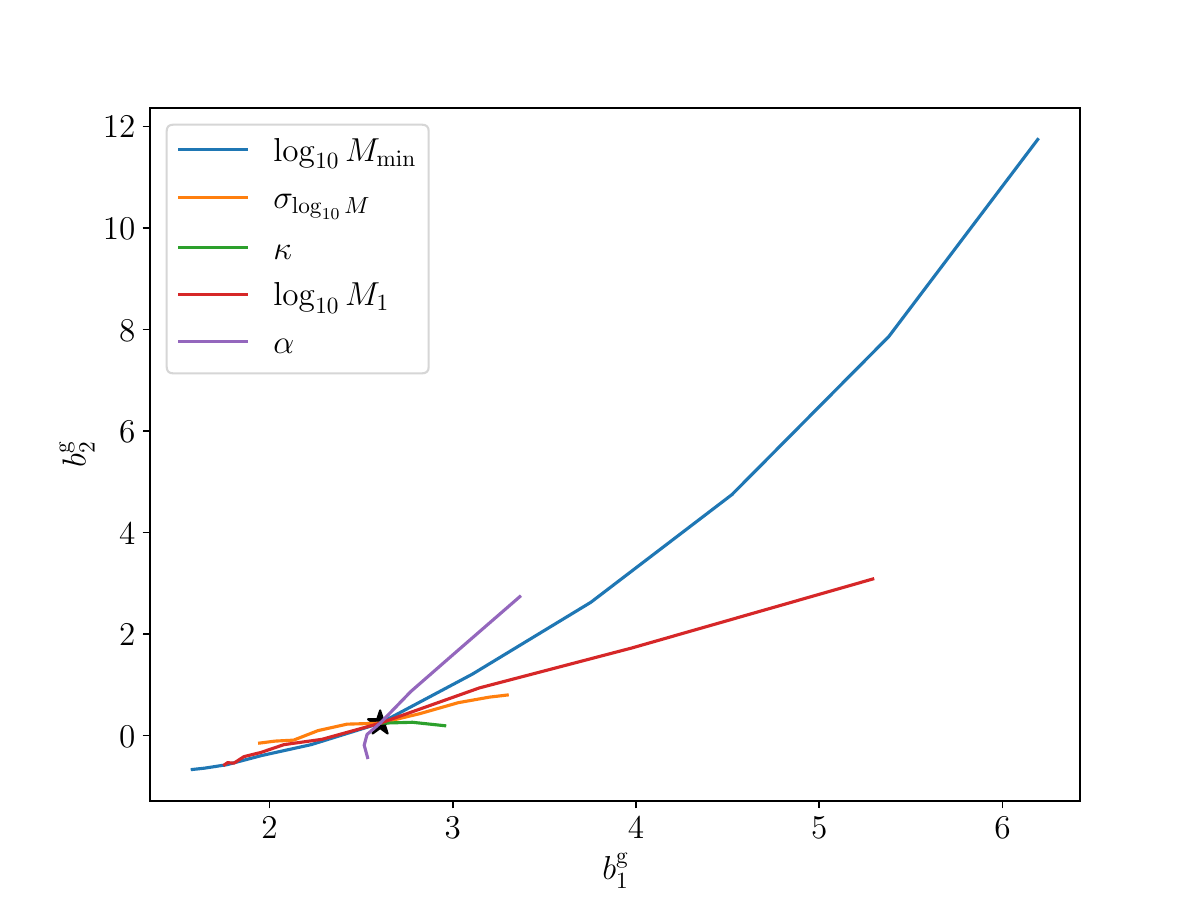}
    \hfill
    \includegraphics[width=0.49\textwidth]{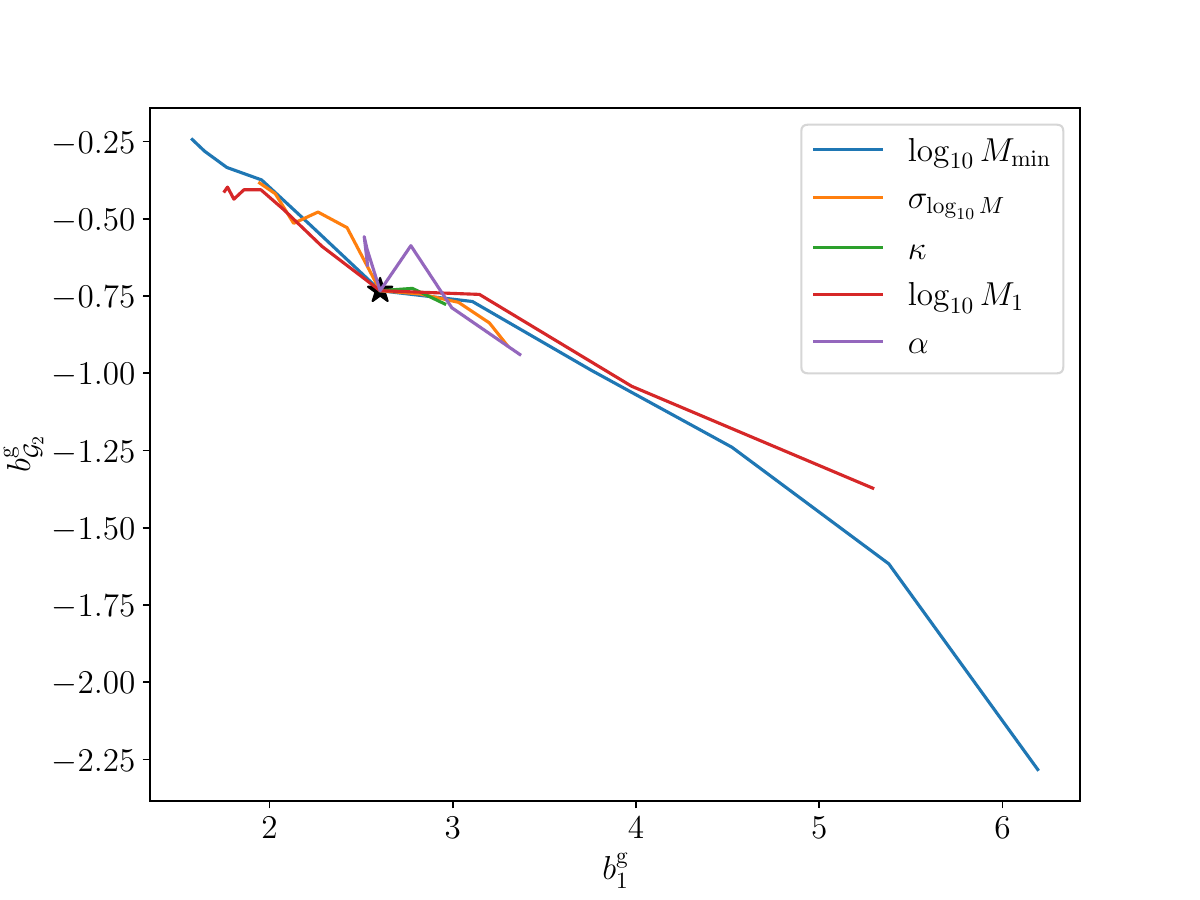}
    \caption{
    The two-dimensional distribution of galaxy bias parameters, $b_1^{\rm g}$, $b_2^{\rm g}$ and $b_{{\cal G}_2}^{\rm g}$ at several redshifts.
    The fiducial set of the HOD parameters is indicated by the star symbol.
    Each different color corresponds to each different HOD parameter.
    }
    \label{fig:bias_HOD}
\end{figure}

Here we focus on the standard HOD prescription where the HOD probability depends only on the halo mass.
%In the following, we present how the galaxy bias parameters vary, depending on each HOD parameter one by one. 
Fig.~\ref{fig:bias_HOD} presents how the galaxy bias parameters vary, depending on each HOD parameter one by one.
The fiducial set of the HOD parameters is $\log_{10}M_{\rm min} = 13.0$, $\sigma_{\log_{10} M} = 0.5$, $\kappa = 1.0$, $\log_{10}M_1 = 13.6$, and $\alpha = 1.0$, which is indicated by the star symbol in Fig.~\ref{fig:bias_HOD} and consistent with both the CMASS and LOWZ galaxies~\cite{Kobayashi:2021oud,Yuan:2022jqf}.
Each colored line corresponds to the variations caused by the selected HOD parameter with other parameters fixed to their fiducial values.
In all cases, the small (large) values of the HOD parameters result in small (large) $b{\rm g}_1$, because all the HOD parameters tend to upweight more massive halos as they take on larger values.

We start by examining the dependence of galaxy bias parameters on $\log_{10}M_{\rm min}$, which is shown by the blue line in Fig.~\ref{fig:bias_HOD}.
It is evident that this parameter induces the largest variations in both $b{\rm g}_1$-$b{\rm g}_2$ and $b{\rm g}_1$-$b{\rm g}_{{\cal G}_2}$, with the HOD parameter range Eq.~\eqref{eq:HOD_range_Mmin}.
This is expected since this parameter controls the halo mass at which galaxies begin to reside in the HOD modeling.
The $\sigma_{\log_{10}M}$ parameter, which controls the softness of the minimum halo mass cut and is shown by the orange line, has a smaller impact on the galaxy bias parameters, 
since it does not change the typical halo mass at which galaxies reside, but only the sharpness of the transition.

The second most important parameter is $\log_{10}{M_1}$, shown by the red line, which determines the overall expected number of satellite galaxies. 
The important observation here is that the slope of the $b_1^{\rm g}$-$b_2^{\rm g}$ relation by $\log_{10}{M_1}$ is different from that by $\log_{10}M_{\rm min}$, while the slopes of the $b_1^g$-$b_{{\cal G}_2}^g$ are similar.
This can be attributed to the fact that satellite galaxies are more sensitive to massive halos than central galaxies, 
and the $b_1^{\rm h}$-$b_2^{\rm h}$ relation is different from the $b_1^{\rm h}$-$b_{{\cal G}_2}^{\rm h}$ relation.
In fact, when $b_1^{\rm h}$ is large, the change in $b_2^{\rm h}$ is also significant, while $b_{{\cal G}_2}^{\rm h}$ remains an almost linear relationship with $b_1^{\rm h}$. 
As a result, when $b_1$ is large, the variation in the number of HOD galaxies induces a non-linear change in $b_2$.
We can interpret the response of the quadratic galaxy bias parameters to $\alpha$ in a similar way.
The $\alpha$ parameter also controls the number of satellite galaxies by changing the slope of the satellite galaxy number as a function of the halo mass.
While larger $\log_{10}{M_1}$ means less satellite galaxies, larger $\alpha$ means more satellite galaxies, leading to the different trend in the $b_1^{\rm g}$-$b_2^{\rm g}$ relation.
The $\kappa$ parameter, which also controls the number of satellite galaxies by changing the minimum halo mass cut for satellite galaxies, has little impact on the galaxy bias parameters.
This is because this parameter has little impact on the number of satellite galaxies, as long as $M_1$ is larger than $M_{\rm min}$, and thus the number of satellite galaxies is determined by $M_1$ and $\alpha$.

In summary, sensitivity of the HOD parameters to the galaxy bias parameters depends both on the HOD parameter itself and the galaxy bias parameter we are interested in, as expected.
In particular, we find that $\log_{10}M_{\rm min}$, $\sigma_{\log_{10}M}$, $M_1$, and $\alpha$ can introduce larger variations compared to $\kappa$, 
and $b_1^{\rm g}$-$b_2^{\rm g}$ relation is more sensitive to the HOD parameters than $b_1^{\rm g}$-$b_{{\cal G}_2}^{\rm g}$ relation.
%The implication we can draw from this result

\subsection{The dependence of the quadratic galaxy bias on redshift}
\label{subsec:bias_redshift}

\begin{figure}[tb]
    \centering
    \includegraphics{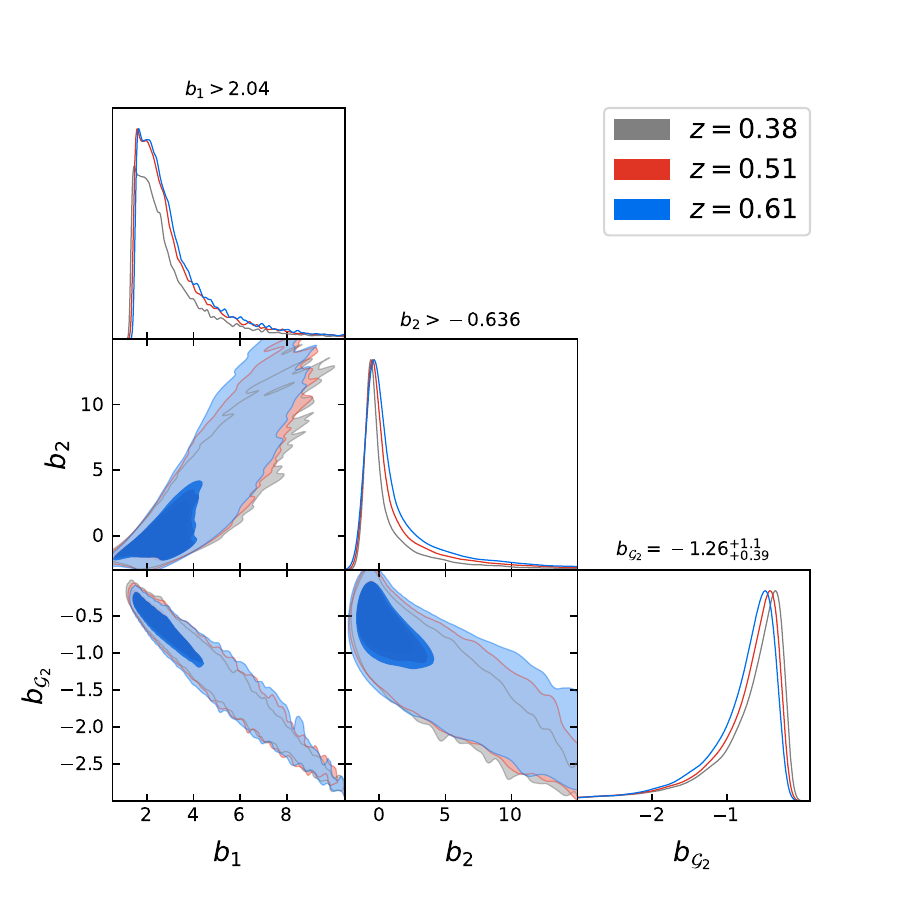}
    \caption{
    The marginal distribution of the quadratic galaxy bias parameters at different redshifts.
    }
    \label{fig:bias_redshift}
\end{figure}
    
Next we discuss the redshift dependence of the galaxy bias parameters.
In Fig.~\ref{fig:bias_redshift}, we show the two-dimensional marginal distributions of quadratic galaxy bias parameters at $z=0.38$, $0.51$, and $0.61$, integrating out all the HOD parameters we vary,
\begin{align}
    {\cal P}({\boldsymbol \theta}_{\rm bias}) = \int \dd {\boldsymbol\theta}_{\rm HOD} {\cal P}({\boldsymbol \theta}_{\rm bias}|{\boldsymbol \theta}_{\rm HOD}) {\cal U}({\boldsymbol \theta}_{\rm HOD}).
\end{align}
In principle, changing redshift can affect the both the halo mass function and the halo bias, 
though it is known that there are the universal relations between $b_1^{\rm h}$, $b_2^{\rm h}$ and $b_{{\cal G}_2}^{\rm h}$ over different redshift and halo mass ranges~\cite{Lazeyras:2015lgp} (see also App.~\ref{app:halo_bias}).
Thus different redshifts just result in the different halo mass functions, which can be absorbed into changing the HOD parameters.
As a result, the distributions of the quadratic galaxy bias parameters are almost the same among different redshifts covered in BOSS survey, shown in Fig.~\ref{fig:bias_redshift}.
Hence, we can conclude that the redshift dependence of the galaxy bias parameters is negligible in the BOSS survey range, as long as the HOD parameters are sampled in the sufficiently large range.

\subsection{The dependence of the quadratic galaxy bias on range of the HOD parameters}
\label{subsec:HOD_range}

\begin{figure}[tb]
    \centering
    \includegraphics{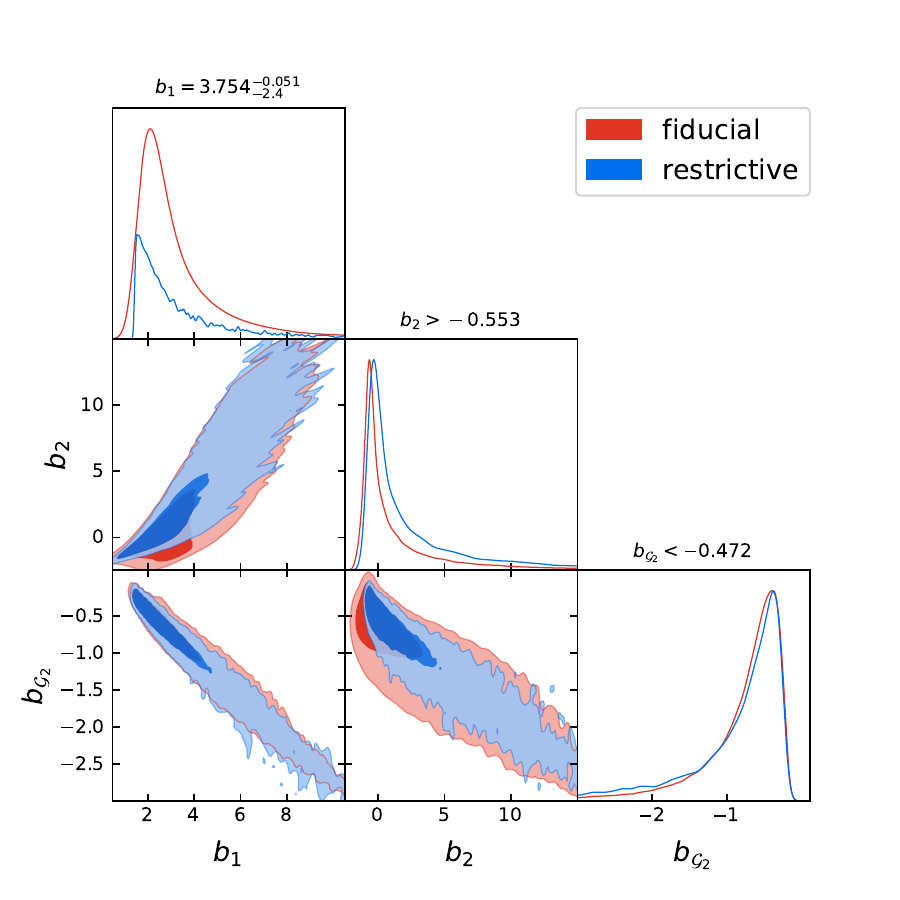}
    \caption{
    The marginal distribution of the quadratic galaxy bias parameters from the different ranges of the HOD parameters.
    The red and blue distributions correspond to our fiducial choice (Eqs.~\eqref{eq:HOD_range_Mmin}-\eqref{eq:HOD_range_alpha}) and restrictive choice (Eqs.~\eqref{eq:HOD_range_Mmin_res}-\eqref{eq:HOD_range_alpha_res}) of the HOD parameters, respectively.
    }
    \label{fig:bias_HOD_restrictive}
\end{figure}

Here we investigate the impact of the range of the HOD parameters on the distributions of the galaxy bias parameters.
In the previous sections, we have shown the distributions of the galaxy bias parameters by varying the HOD parameters in the range Eqs.~\eqref{eq:HOD_range_Mmin}-\eqref{eq:HOD_range_alpha}.
In this section, we consider the more restrictive range of the HOD parameters, which is given by
\begin{align}
    \log_{10}M_{\rm min} \in & [12.4, 14.2]\,\,,
    \label{eq:HOD_range_Mmin_res}
    \\
    \sigma_{\log_{10}M} \in & [0.1, 1.0]\,\,,
    \label{eq:HOD_range_sigmaM_res}
    \\
    \kappa \in & [0.1, 1.0]\,\,,
    \label{eq:HOD_range_kappa_res}
    \\
    \log_{10} M_1 \in & [13.2, 14.4]\,\,,
    \label{eq:HOD_range_M1_res}
    \\
    \alpha \in & [0.8, 1.6]\,\,.
    \label{eq:HOD_range_alpha_res}
\end{align}
We restrict the range of $\log_{10} M_1$ and $\alpha$, which are responsible for the number of satellite galaxies, 
while we keep the same ranges for $\log_{10}M_{\rm min}$ and $\sigma_{\log_{10}M}$, which are responsible for the number of host galaxies, 
Note that since $\kappa$ has little impact on the galaxy bias parameters, as we have seen in Sec~\ref{subsec:bias_HOD_fiducial}, we keep the same range for $\kappa$.

The result is shown in Fig.~\ref{fig:bias_HOD_restrictive}.
As expected, the restrictive range of the HOD parameters leads to the more restrictive distributions of the galaxy bias parameters.
However, it is worth noting that the marginal distribution of $b_1^{\rm g}$-$b_{{\cal G}_2}^{\rm g}$ remains the almost same, while that of $b_1^{\rm g}$-$b_2^{\rm g}$ is more sensitive to the range of the HOD parameters.
This can be understood from the fact that the $b_1^{\rm h}$-$b_2^{\rm h}$ relation is quite nonlinear at larger $b_1^{\rm h}$, as we have discussed in Sec.~\ref{subsec:bias_HOD_fiducial}.
This result implies that the choice of the range of the HOD parameters can change the distributions of $b_2^{\rm g}$ significantly, while the distributions of $b_{{\cal G}_2}^{\rm g}$ are less affected.

\subsection{The dependence of the quadratic galaxy bias on the HOD modeling and the assembly bias}
\label{subsec:HOD_modeling}

\begin{figure}[tb]
    \centering
    \includegraphics{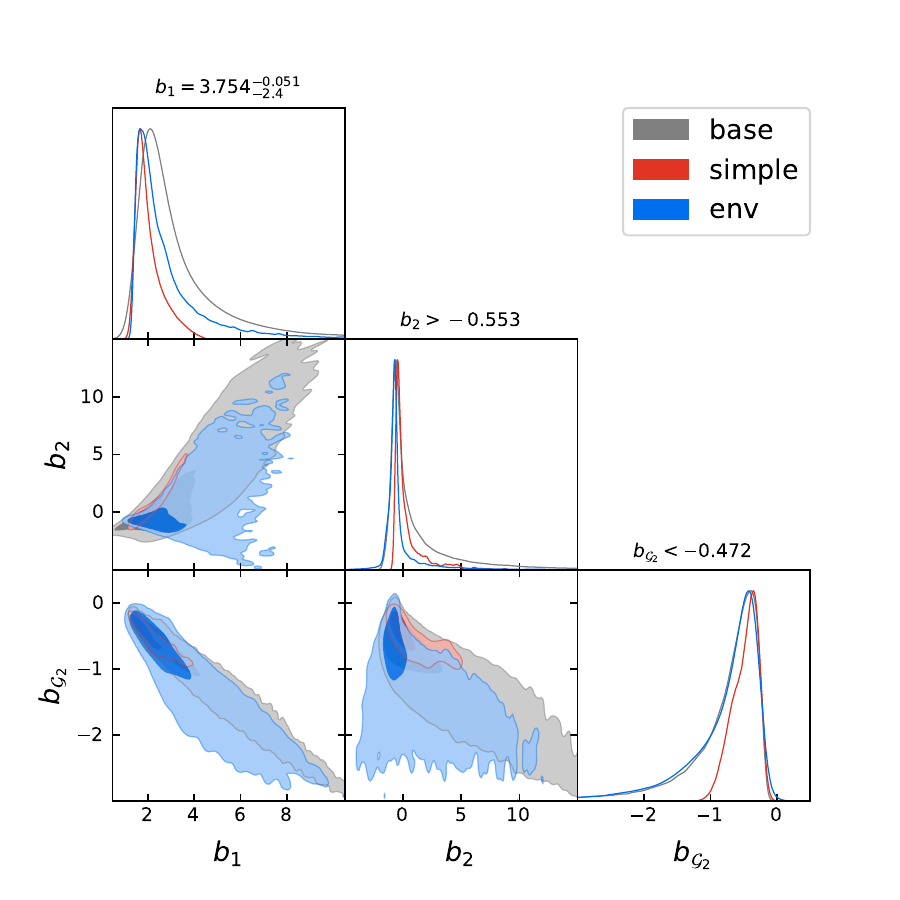}
    \caption{
    The marginal distribution of the quadratic galaxy bias parameters from different HOD models.
    The grey, red, and blue distributions correspond to the standard HOD, the simplified HOD, and the standard HOD with concentration, respectively.
    }
    \label{fig:bias_HODs}
\end{figure}

So far we only employ the standard HOD model, where the galaxy-halo connection is assumed to be solely determined by the host halo mass with five paramters.
However, several studies show that (1) the halo bias has some dependence other than its mass, and (2) the halo-galaxy connection should involve some complexities beyond the host halo mass~\cite{Gao:2005ca,Wechsler:2005gb,Jing:2006ey,Dalal:2008zd,Zentner:2013pha,Lazeyras:2016xfh,Lazeyras:2021dar}.
In order to investigate the impact of the different HOD modeling and the assembly bias on the distributions of the galaxy bias parameters, here we show the results of the simplified HOD and the standard HOD with concentration, which are described in Sec.~\ref{subsec:HOD}.

Fig.~\ref{fig:bias_HODs} shows the two-dimensional marginal distribution of the quadratic galaxy bias parameters for the simplified HOD (the red) and the standard HOD with concentration  (the blue) along with the standard HOD (the grey).
The simplified HOD model leads to more restrictive distributions of the galaxy bias parameters compared to the standard HOD, which is expected since the simplified HOD model has fewer parameters.
In addition, the coverage of the galaxy bias parameters by the simplified HOD model are within the range of the standard HOD model.
On the other hand, the standard HOD with concentration model shows different distributions of the galaxy bias parameters compared to the standard HOD ones, while they are partially overlapped.
Overall, when considering the concentration dependence in the way explained in Sec.~\ref{subsec:HOD}, the distributions of galaxy bias parameters extend to smaller values across all the galaxy bias parameters we examine.
This is consistent with the fact that the halos with higher concentrations tend to have smaller $b_1^{\rm h}$, $b_2^{\rm h}$, and $b_{{\cal G}_2}^{\rm h}$~\cite{Lazeyras:2021dar}, 
since in this mock we populate galaxies into halos with higher concentrations preferentially.
This result implies that the distributions of the galaxy bias parameters can be affected by the assembly bias significantly, which is not included in the standard HOD model 
and one must be careful about how the effects beyond the standard HOD model is taken into account when employing the HOD-informed prior on the galaxy bias parameters.

\subsection{Possible prior on quadratic galaxy bias}
\label{subsec:bias_prior}

\begin{figure}[tb]
    \centering
    \includegraphics[width=0.49\textwidth]{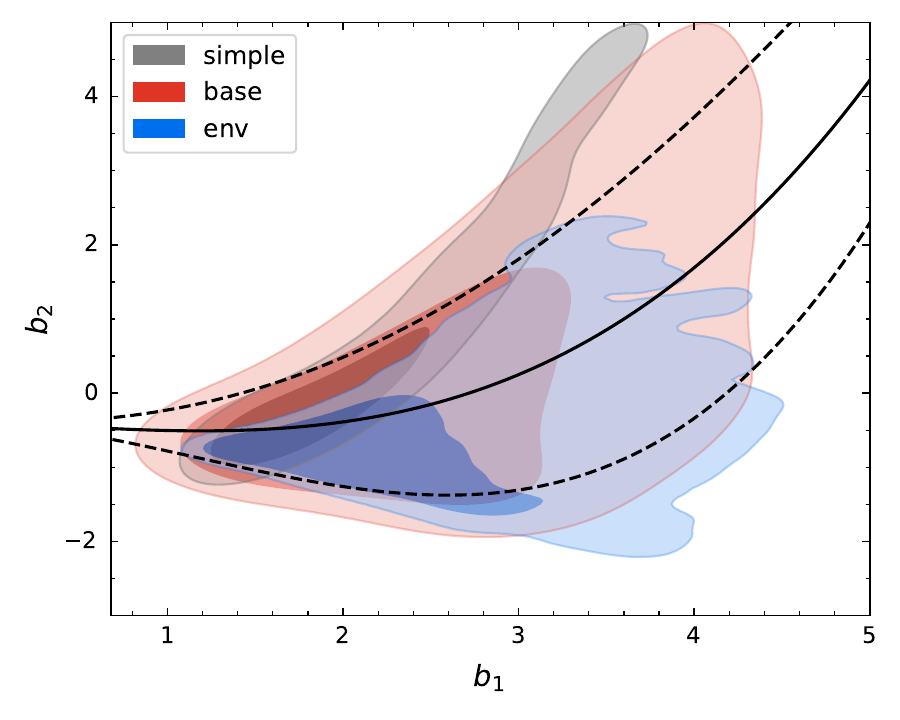}
    \hfill
    \includegraphics[width=0.49\textwidth]{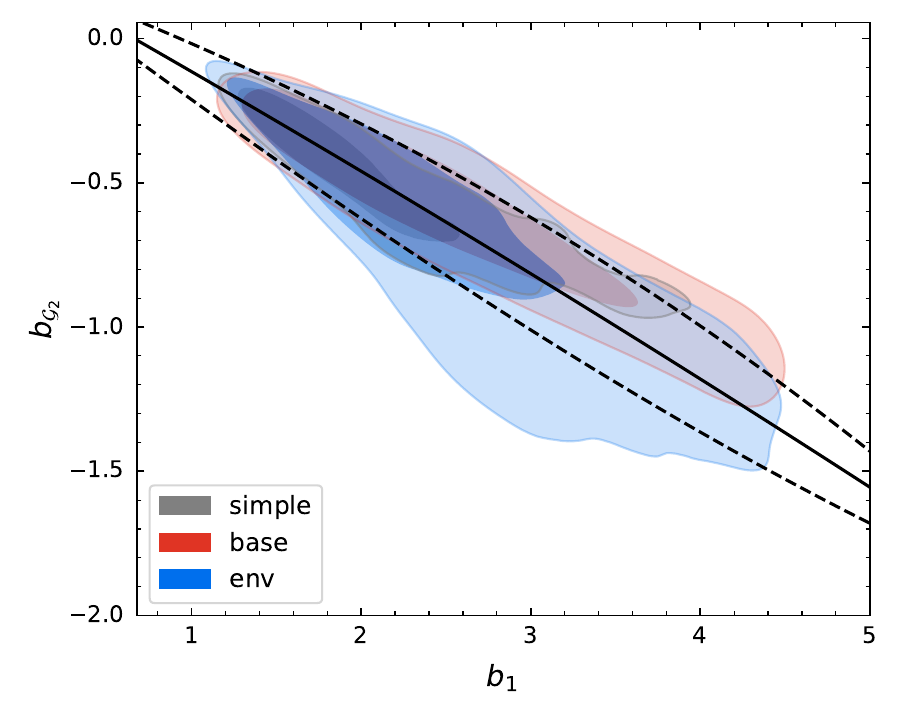}
    \caption{
    The two-dimensional marginal distributions of galaxy bias parameters, $b_1^{\rm g}$-$b_2^{\rm g}$ and $b_1^{\rm g}$-$b_{{\cal G}_2}^{\rm g}$, conditioned to $b_1 \in [0.0, 5.0]$.
    The best fit lines are shown by the solid lines and the $1\sigma$ is shown by the dashed lines, which are obtained in $b_1 \in [1.0, 4.0]$. 
    }
    \label{fig:bias_HOD_prior}
\end{figure}

In this subsection, we discuss the possible prior on the galaxy bias parameters, focusing on the BOSS-like galaxy samples.
Although the distributions of the galaxy bias parameters depend on the assumed galaxy-halo connection, 
the results in the previous sections suggest that the quadratic galaxy bias parameters are correlated with the linear galaxy bias parameter
and thus it makes sense to consider the prior on the quadratic galaxy bias parameters conditioned to the linear galaxy bias parameter.

Fig.~\ref{fig:bias_HOD_prior} shows the two-dimensional marginal distributions of the galaxy bias parameters, $b_1^{\rm g}$-$b_2^{\rm g}$ and $b_1^{\rm g}$-$b_{{\cal G}_2}^{\rm g}$, conditioned to $b_1 \in [0.0, 5.0]$.
Note that here we use the our fiducial choice of the HOD parameter range, Eqs.~\eqref{eq:HOD_range_Mmin}-\eqref{eq:HOD_range_alpha} and combine the samples from three different redshifts to get more samples, given that the different redshifts barely affect the marginal distributions of the galaxy bias parameters as discussed in Sec.~\ref{subsec:bias_redshift}.
In principle the obtained distribution of the galaxy bias parameters itself can be used as the priors on the galaxy bias parameters. 
However, it is a non-trivial task to get a functional form of this distribution in order to generate new samples under this distribution.
One method to get around this problem is to use normalizing flows (see Refs.~\cite{Ivanov:2024xgb,Ivanov:2024hgq,Zhang:2024thl}), but this is beyond the scope of this paper.
Instead, in this work, we just obtain simple polynomial fitting formulas for $b_2^{\rm g}$ and $b_{{\cal G}_2}^{\rm g}$ as functions of $b_1^{\rm g}$ for convenience, 
although this is not an optimal use of the obtained probability distribution. 
Specifically, here we provide the mean relations between $b_1^{\rm g}$-$b_2^{\rm g}$ and $b_1^{\rm g}$-$b_{{\cal G}_2}^{\rm g}$ as well as the $1\sigma$ deviations from these mean relations 
as functions of $b_1^{\rm g}$ that approximate the distributions in the 2D planes:

\begin{align}
    b_2^{\rm g}(b_1^{\rm g}) & =  -0.38 -0.15 b_1^{\rm g} - 0.021 (b_1^{\rm g})^2 + 0.047 (b_1^{\rm g})^3,
    \\
    \sigma(b_2^{\rm g})(b_1^{\rm g}) & =  0.06 b_1^{\rm g} + 0.24 (b_1^{\rm g})^2 - 0.02 (b_1^{\rm g})^3 - 0.003 (b_1^{\rm g})^4,
    \\
    b_{{\cal G}_2}^{\rm g}(b_1^{\rm g}) & =  0.22 - 0.33 b_1^{\rm g} - 0.005 (b_1^{\rm g})^2,
    \\
    \sigma(b_{{\cal G}_2}^{\rm g})(b_1^{\rm g}) & =  0.11  b_1^{\rm g} - 0.012 (b_1^{\rm g})^2 - 0.001 (b_1^{\rm g})^3 .
\end{align}
This priors are already used in Ref.~\cite{Cabass:2024wob} to get better constraints on massive particles during inflation from the BOSS data.
%We note that the $b_1^{\rm g}$-$b_{{\cal G}_2}^{\rm g}$ relation is stable against the different choice of the HOD model, while $b_1^{\rm g}$

\section{Discussion}
\label{sec:discussion}

We present the dependence of the quadratic galaxy bias parameters on the HOD parameters, redshift, and the different HOD models.
While {\tt ABACUSHOD} \cite{Yuan:2022jqf}, used in Refs.~\cite{Ivanov:2024xgb,Ivanov:2024hgq,Zhang:2024thl}, takes into account the dependence on the local density, 
our study has demonstrated that incorporating different assembly bias model with halo concentration could alter the priors on galaxy bias parameters. 
This suggests that it is worthwhile to explore the responses to other forms of assembly bias, such as those related to halo formation history and ellipticity, to fully comprehend their impact on galaxy bias parameters.

Related to this direction, 
while we employ the HOD model to populate galaxies into halos, investigating the outcomes using alternative methods is equally important. 
The SubHalo Abundance Matching (SHAM) technique is widely used for assigning galaxies to halos based on their subhalo properties~\cite{Vale:2004yt,Conroy:2005aq,Behroozi:2010rx}. 
Examining how the marginal distributions of the galaxy bias parameters differ when using SHAM could provide valuable insights into the dependence of galaxy bias parameters on the galaxy-halo assignment method. 
This comparison would help to assess the robustness of the prior of the galaxy bias parameters and determine whether it is sensitive to the specific population technique employed.

Our analysis revealed that the scatter in the relationship between the linear bias $b_1^{\rm g}$ and the tidal bias $b_{{\cal G}_2}^{\rm g}$ is not significantly larger than that between $b_1^{\rm g}$ and $b_2^{\rm g}$ across different HOD models.
This observation suggests the existence of an optimal basis of bias parameter combinations that are less sensitive to the details of the HOD modeling. 
Identifying such a basis could be beneficial for obtaining the robust priors on the galaxy bias parameters.

Finally, in this work we measure the bias parameters using the quadratic field method, which benefits from reduced sample variance by utilizing cross-correlations between the galaxy density field and quadratic combinations of the initial density field. 
Although we show the consistency between our method and the separate universe method in App~\ref{app:halo_bias},
it is crucial to explicitly check the consistency of our results with those obtained from other methods, such as the power spectrum and bispectrum analyses~\cite{Nishimichi:2020tvu,Eggemeier:2021cam, Ivanov:2021kcd, Zhang:2024thl} or field-level comparison based on the Lagrangian description~\cite{Schmittfull:2018yuk,Ivanov:2024xgb,Ivanov:2024hgq}. 
%Cross-validating with independent techniques would strengthen the reliability of our bias measurements and ensure that our conclusions are not artifacts of the specific method used.
We will leave these for the future work.

\section*{Acknowledgement}
We thank  Stephen Chen,  Misha Ivanov, Yosuke Kobayashi, and Marko Simonovi\'c for useful discussions.
KA is supported by Fostering Joint International Research (B) under Contract No.21KK0050 and the Japan Society for the Promotion of Science (JSPS) KAKENHI
Grant No.JP24K17056.
Numerical computation was carried out on Cray XC50 at Center for Computational Astrophysics, National Astronomical Observatory of Japan.

\appendix

\section{Measurement of the halo bias parameters}
\label{app:halo_bias}
In this Appendix, we present the precise measurement of the quadratic halo biases, ($\tilde{b}_2^{\rm h}$, $b_{K^2}^{\rm h}$).\footnote{Note that here we present ($\tilde{b}_2^{\rm h}$, $b_{K^2}^{\rm h}$), not ($b_2^{\rm h}$, $b_{{\cal G}_2}^{\rm h}$) shown in the main text to make the comparison with the literature easier.}
These halo bias parameters have already been measured accurately in several ways in the literature, including the use of the bispectrum, the quadratic field method, and the separate universe simulation~\cite{Saito:2014qha,Schmittfull:2014tca,Lazeyras:2017hxw,Abidi:2018eyd,Eggemeier:2021cam}.

\begin{figure}[tb]
\centering
\includegraphics[width=0.48\textwidth]{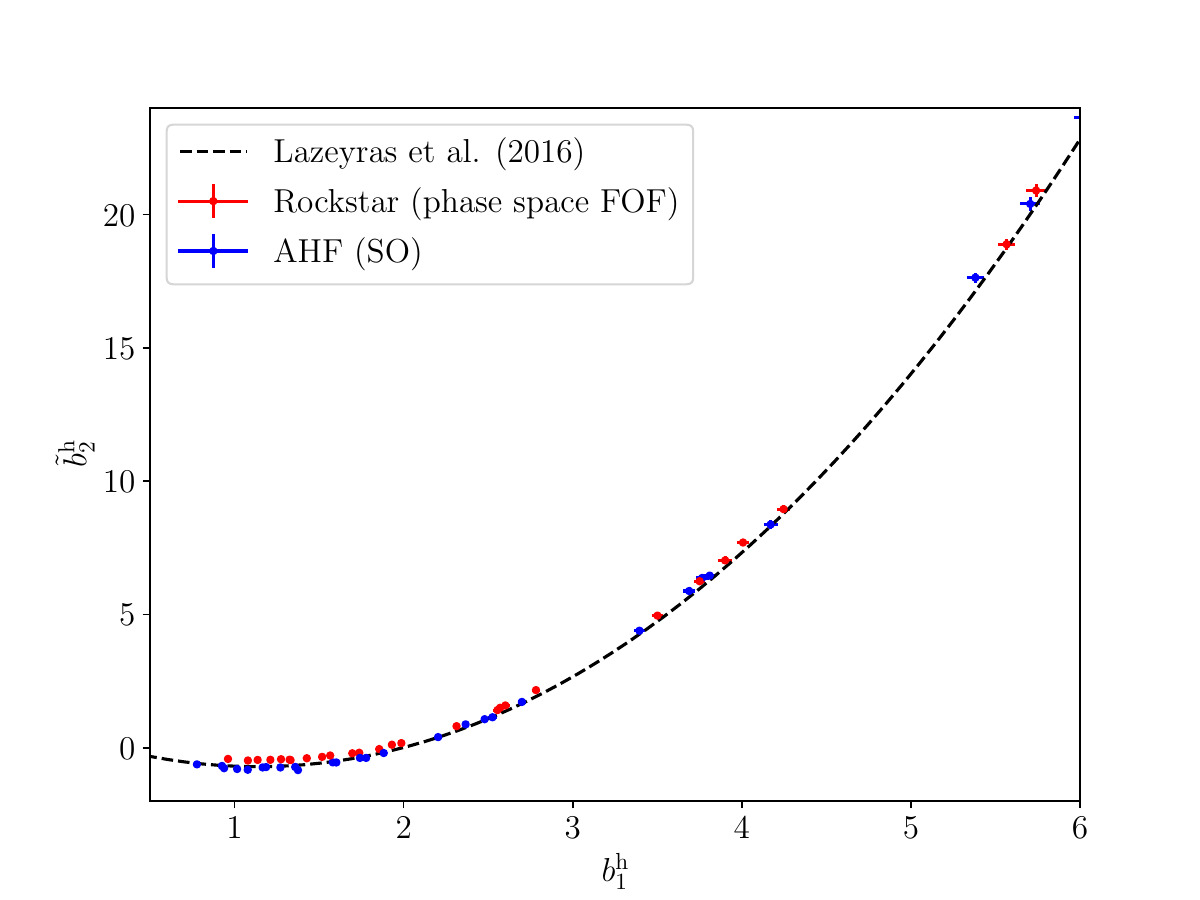}
\hfill
\includegraphics[width=0.495\textwidth]{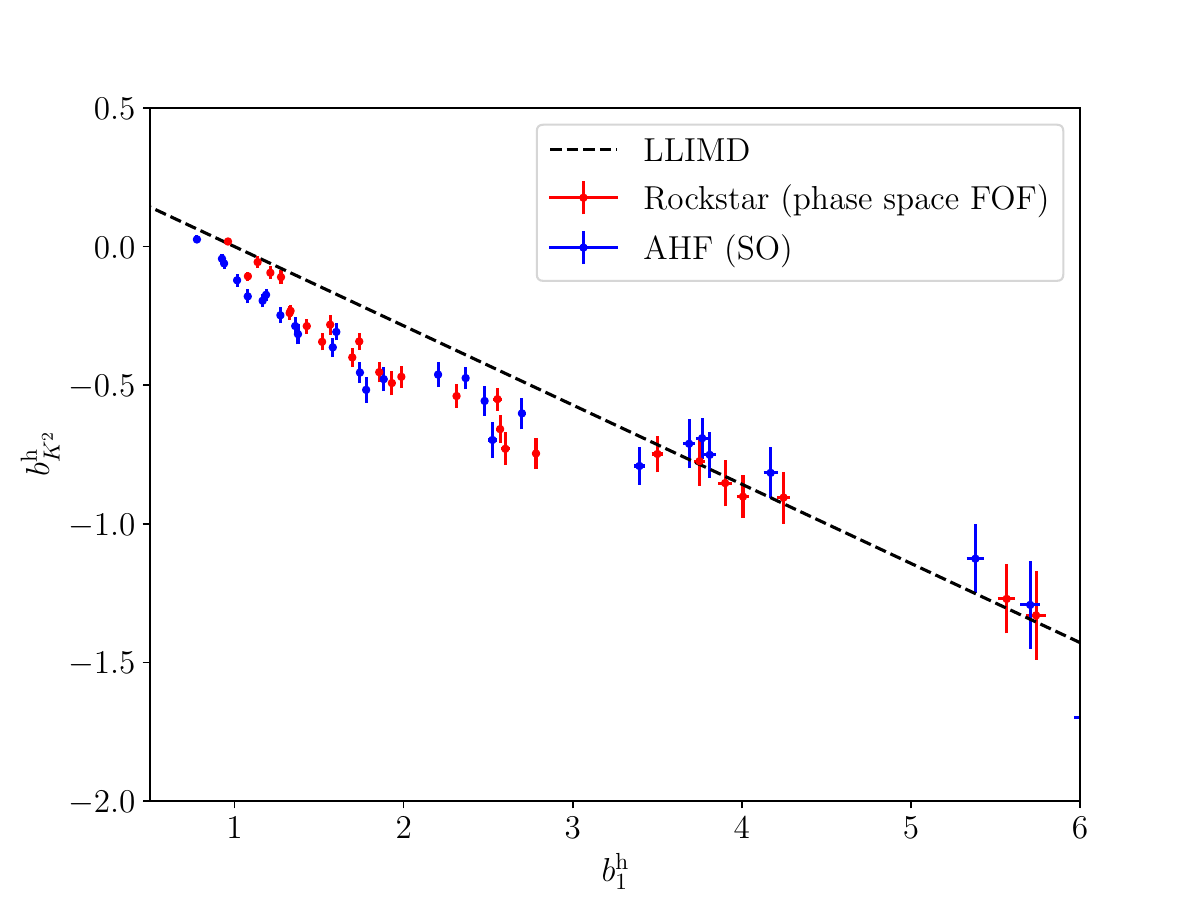}
\caption{
The quadratic halo bias parameters, $\tilde{b}_2^{\rm h}$ and $b_{K^2}^{\rm h}$, as a function of the linear bias parameter $b_1^{\rm h}$ for various mass halos at various redshifts.
In the left panel we also plot the fitting formula of $b_2^{\rm h}$ from Ref.~\cite{Lazeyras:2015lgp} by the dashed curve.
The dashed line in the right panel, on the other hand, shows the Lagrangian local-in-matter-density (LLIMD) ansatz (or also known as the coevolution ansatz), $b_{K^2}^{\rm h} = -\frac27 (b_1^{\rm h}-1)$.
}
\label{fig:halo_bias}
\end{figure}

Fig.~\ref{fig:halo_bias} shows our measurements of $\tilde{b}_2^{\rm h}$ and $b_{K^2}^{\rm h}$ as a function of $b_1^{\rm h}$ for the halo mass from $10^{12} M_\odot/h$ to $10^{15} M_\odot/h$ at various redshifts.
Since Ref.~\cite{Lazeyras:2015lgp}, using the separate universe simulation method, provides the most precise measurement for the local halo bias parameters and obtain the fitting formula for $\tilde{b}_2^{\rm h}$ as a function of $b_1^{\rm h}$,
we also show this fitting by the blue solid curve in the $b_1^{\rm h}$-$\tilde{b}_2^{\rm h}$ plot (the left panel).
Our results are in excellent agreement with the fitting while the cosmology, the $N$-body simulation details, the halo finders, and the methods to measure the bias parameters are completely different among ours and Ref.~\cite{Lazeyras:2015lgp}.
The comparison and agreement with this previous work provides the consistency check of our method.

In the $b_1^{\rm h}$-$b_{K^2}^{\rm h}$ plot (the right panel) the dashed line corresponds to the Lagrangian local-in-matter-density (LLIMD) (or coevolution) prediction for the tidal bias: $b_{K^2}^{\rm h} = -\frac27(b_1^{\rm h}-1)$~\cite{Chan:2012jj,Baldauf:2012hs}.
We find the small deviations from the LLIMD prediction, which is consistent with the previous works~\cite{Lazeyras:2017hxw,Abidi:2018eyd,Barreira:2021ukk,Akitsu:2022lkl}, except for Ref.~\cite{Modi:2016dah}.

\bibliography{references}
\end{document}